# TITLE: DC Signature of snap-through bi-stability in carbon nanotube resonators

AUTHORS: Sharon Rechnitz, Tal Tabachnik, Shlomo Shlafman, Michael Shlafman and Yuval E. Yaish[*].

Andrew and Erna Viterbi Faculty of Electrical and Computer Engineering, Technion, Haifa, Israel.

ABSTRACT: Bi-stable arched beams exhibiting Euler-Bernoulli snap-through buckling are vastly used as electronic devices in various applications, such as memory devices, energy harvesters, sensors, and actuators. Recently, we reported the realization of the smallest bi-stable resonator to date, in the form of a buckled suspended carbon nanotube (CNT), which exhibits a unique three-dimensional snap-through transition and an extremely large change in frequency as a result. In this article, we address a unique characteristic of these devices, in which a significant change in the DC conductance is also observed at the mechanical snap-through transition. After verifying that the change in the CNT tension due to the "jump" cannot account for the conductance difference measured, we attribute the conductance "jump" to the change in capacitance as a result of the snap-through buckling. However, we show that quantitative analysis of this phenomenon is not at all trivial, and is enabled only due to our ability to predict the exact CNT shape before and after the transition. Understanding this mechanism enables a fast characterization of fabricated devices and improves our understanding of their behavior, key in developing this technology further and better design. As an example, we show how the hysteretic trait of this phenomenon is indicative of the ability to achieve static latching, useful for RF switches, bi-stable relays, and memory devices.

TEXT:

An important feature for a micro/nano-electromechanical system (MEMS/NEMS) is the ability to tune the state of the moveable object by an external parameter, most widely used is by electric fields. A common prototype is based on a conductive beam electrostatically actuated by applying a voltage difference between a nearby electrode and the movable beam[1]. The applied voltage can alter the static position of the beam as well as to excite its resonance modes[2]. Numerous studies have examined the properties of such suspended beams under static and dynamic forces, including their linear and nonlinear behavior due to the combination of mechanical restoring force and electric field[1,3–6]. Specifically, an arch shaped beam which undergoes Euler-Bernoulli buckling instability is commonly used as a bi-stable device[7–9]. In such initially curved beam clamped at both ends and actuated by electrostatic force, a non-monotonous stiffness-deflection characteristic is found, and snap-through (ST) buckling phenomenon can be observed[10]. Under these circumstances, the mechanical constraint limits the beam movement, making the system stiffer after the ST transition, and an additional stable equilibrium appears. Such structures are suitable for various applications including sensors[11], memory devices[12,13], actuators[14,15], filters[16], micro-valves[17], and buckling-induces smart applications[18].

Recently, we reported the first realization of ST bi-stability in CNT resonators, which results from initial upward buckling of the CNT[19]. We also reported the evidence of bi-stability in a DC conductance measurement of the device, in the form of a discontinuity ("jump") of the measured current (Fig. S1 in Ref. 19). In this study, we present extensive data in which this discontinuity was observed. We analyze and explain the origin of this phenomenon, which turns out to be more complex than initially anticipated.

**Results and Discussion**

Fig. 1a,b presents a resonance frequency measurement of a typical bi-stable CNT resonator for downward and upward DC gate sweep, respectively. The CNT is initially curved upward[19]. As the gate voltage (absolute value) increases, the CNT is attracted toward the local gate, and the resonance frequency decreases due to compression. At a certain point ($V_{gDC,ST}$ = 2.77 V for upward sweep or $V_{gDC,ST}$ = -2.63 V for downward sweep), the CNT cannot compress any further and "jumps" to a downward configuration, a transition known as snap-through buckling. This mechanical transition results in a "jump" in the resonance frequency, marked by the yellow arrows in Fig. 1a,b. After the transition, increasing the gate voltages further stretches the CNT and the resonance frequency increases. When the voltage is swept back to zero, the resonance frequency decreases as the stretching is gradually reduced, until reaching a second minimum, at which a snap-back (or "release") transition occurs, also marked by the vertical yellow arrows ($V_{gDC,R}$ = -2.47 V for upward sweep or $V_{gDC,R}$ = 2.62 V for downward sweep). The difference between the gate voltages at which the ST (downward "jump") and release (upward "jump") transitions occur creates a hysteresis window. As an example, for positive gate voltages - $\Delta V_{hyst} = V_{gDC,ST} - V_{gDC,R}$ = 0.15 V. We shall clarify that the actual CNT static motion is more complex than described in this simplified explanation, since an out-of-plane deflection also evolves[19]. We developed a theoretical model that allows us to predict the exact CNT shape at every static load[19,27], but ultimately, the snap-through "jump" represents a sudden transition from upward to downward configuration. Surprisingly, we observe a discontinuity in the DC transfer characteristic curves of bi-stable CNT resonators, at the same load as the snap-through transition occurs (Fig. 1c).

Mechanical vibrations of a suspended CNT will usually have no effect on its DC conductance. However, since the Euler-Bernoulli snap-through transition results in a relatively large mechanical motion, it is also evident in a DC conductance measurement. Fig. 2 presents data of similar DC "jumps" obtained from 20 different bi-stable CNT resonators. Each shape and color represent a device, where a square/diamond frame represents whether the "jump" occurs for P/N-type CNT, respectively. δI is the DC current difference measured before and after the "jump", plotted as a function of the DC measurement slope $\partial I/\partial V_G$. It can be observed that the majority of the "jumps" are positive (i.e., an increase in the conductance due to the ST transition), but occasionally we detect a negative "jump" (a decrease in the conductance). Note that negative "jumps" occur only for small $\partial I/\partial V_G$ values, meaning at saturation where there is nearly no conductance modulation due to change in the gate voltage, implying that the device conductance is mainly restricted due to contact resistance.

In the following discussion we attempt to explain the ST DC "jumps". The discussion will follow a single example, based on the device presented in Fig. 1 (Device I), but we also present the results obtained in a similar manner to two other devices in the Supplementary Information, arriving at the same conclusions.

**Naïve capacitance model.** Intuitively, we attribute the change in conductance to the change in capacitance due to the large mechanical snap-through transition. As a result, the charge induced upon the CNT by the local gate will change and hence the current:

$$\delta I = \frac{\partial I}{\partial q}\delta q = \frac{\partial I}{\partial(V_g^{DC}C_g)}\delta(V_g^{DC}C_g) = \frac{\partial I}{C_g \partial V_g^{DC}}V_g^{DC}\delta C_g = \frac{\partial I}{\partial V_g^{DC}}V_g^{DC}\frac{\partial C_g}{C_g \partial z}\delta z \quad \text{(Eq. 1)}$$

where q is the charge induced upon the CNT, $C_g$ is the capacitance between the CNT and the local gate, and the derivative is taken at constant gate voltage. z represents the in-plane CNT deflection, and δz=|z_{beforeST} - z_{afterST}| is the change of the CNT displacement due to the ST transition. The physical parameters of the device are detailed in Table 1.

Taking the capacitance of a wire parallel to plane, and assuming $\frac{(g_0+z)^2}{r^2} \ll 1$ (where z is the CNT deflection along the z axis), we receive[27]: $\frac{\partial C_g}{C_g \partial z} = \left((g_0 + z) \ln \frac{2(g_0+z)}{r}\right)^{-1}$. Assuming z<<$g_0$, we obtain the relation:

$$\delta I = \left(\frac{\partial I}{\partial V_g^{DC}} V_g^{DC} \frac{1}{g_0 \ln \frac{2g_0}{r}}\right) \delta z \quad \text{(Eq. 2)}$$

Hence, we should be able to estimate the mechanical jump δz from the jump in the conductance. However, if we substitute all the parameters extracted from Fig. 1c into Eq. 2, we get an estimation of $\delta z \approx 151 nm \approx g_0$, which is inconsistent with our assumption.

Fortunately, our theoretical model[19] for the resonance frequency modes versus the gate voltage (Fig. 3a) allows us to determine the exact shape and location of the CNT for any static load, and specifically before and after the ST transition (Fig. 3b). Please note that our convention is such that the positive z axis points downwards. For capacitance calculations, only the in-plane (z) component is relevant (Fig. 3c). Examining the in-plane CNT shape raises the question of how δz should even be estimated. Before we answer this question, we shall notice that even if we consider the maximum deflection $\delta z_{max} = max(z_{down}(x)) - min(z_{up}(x)) = 52.9 nm$, it can only account for a "jump" of $\delta I = 1.45 nA$, only a quarter of the experimental $\delta I_{exp} = 4.77 nA$. Therefore, taking a more realistic δz of, for example, the average displacement (i.e., $\delta z = avg(z_{down}(x)) - avg(z_{up}(x))$), will also fall short. In addition, the capacitance model cannot account for the negative "jumps" (a decrease in the conductance as a result of the ST transition) occasionally detected (Fig. 2 and Fig. S1).

**Modulation of the band gap as a result of strain.** The mechanical ST transition involves also a sudden change in the axial tension along the tube, evidenced also as the "jump" in frequency (Fig. 1a,b). Strain can both enlarge as well as reduce the band gap[20-22], which can potentially explain both an increase as well as a decrease in the conductance, as observed in the data. Since we know the exact shape of the CNT for every static load, we can calculate the tension along the tube as well as the strain, $\epsilon = \frac{\Delta L}{L_0} = \frac{T}{EA}$, where T is the axial tension, E is the CNT Young's modulus, A is the cross section area, $L_0$ is the relaxed CNT length (if no tensile forces are applied), and ΔL is the change in the CNT length due to strain. For the CNT in Fig. 3 we extract $T_{up} = 0.1339 \, nN$, the equivalent of $\epsilon_{up} = 2.8838 \cdot 10^{-5}$, and $T_{down} = 0.1276 \, nN$, the equivalent of $\epsilon_{down} = 2.7476 \cdot 10^{-5}$, meaning that the change in strain as a result of the ST transition is $\delta\epsilon = 1.3623 \cdot 10^{-6}$. The band gap dependence on the strain[20] is given by $\delta E_g = \pm(3t_0(1+\nu)\cos(3\phi)) \cdot \delta\epsilon$, where $t_0$ is the tight-binding overlap integral, ν is the Poisson ratio, and ϕ is the CNT chiral angle. As an upper bound estimation, we substitute typical values from Refs. [21,22] ($t_0 \approx 2.7$ and $\nu \approx 0.2$) and assume armchair CNT ($\cos(3\phi)$=1). This results in a change in the band gap of $|\delta E_g| \approx 1.3242 \cdot 10^{-5} eV$. How should this translate into a change in the CNT conductance?

In order to estimate the resulting expected change in current (δI), we need to know the band gap of the device. For this purpose, we have developed a Drude-based theoretical model for the current gate dependence for single-wall CNT, as depicted in Fig. 1. We begin with the basic relation:

$$I = G_T V_{DS} = (G_T^n + G_T^p) V_{DS} \quad \text{(Eq. 3)}$$

where $V_{DS}$ is the voltage difference between the two contacts, and $G_T$ is the total conductance, which is the sum of the electrons' conductance and the holes' conductance, each comprised of the intrinsic CNT conductance in series with the contact resistance:

$$\left(G_T^{n/p}\right)^{-1} = \frac{L}{\sigma_{n/p}} + R_{Cn/p} \quad \text{(Eq. 4)}$$

where $\sigma_{n/p}$ and $R_{Cn/p}$ are the CNT conductivity and the contacts resistance for electrons/holes, respectively. The CNT conductivity is estimated according to the Drude model:

$$\sigma_{n/p} = e \cdot \mu_{n/p} \cdot n(V_g)/p(V_g) \quad \text{(Eq. 5)}$$

where $\mu_{n/p}$ is the mobility and n/p is the number of negative/positive charges per length, which is a function of the gate voltage, given by:

$$n = \int_\Delta^\infty \nu(\varepsilon) f_{FD}(\varepsilon - \tilde{\mu}) d\varepsilon$$

$$p = \int_{-\infty}^{-\Delta} \nu(\varepsilon) f_{FD}(\tilde{\mu} - \varepsilon) d\varepsilon \quad \text{(Eq. 6)}$$

where $f_{FD}$ is the Fermi-Dirac distribution, $\tilde{\mu}(V_g)$ is the chemical potential, $\varepsilon$ is energy, $\Delta$ is half of the band gap ($\Delta = E_{gap}/2$), and $\nu$ is the density of states given by:

$$\nu(\varepsilon) = \frac{4}{\pi \hbar v_F} \frac{\varepsilon}{\sqrt{\varepsilon^2 - \Delta^2}}$$

where $v_F$ is the Fermi velocity.

When non-zero voltage is applied to the gate, the CNT electric potential ($\varphi_{NT}$) is given by:

$$e\varphi_{NT} = \frac{e^2(n-p)}{C_T} - \frac{eC_g}{C_T} V_g \quad \text{(Eq. 7)}$$

where $C_T, C_g$ are the total and gate capacitances per length, respectively, and $\alpha_g = \frac{C_g}{C_T}$ is the gate efficiency factor, which can be extracted from our fitting to the resonance measurement (more details in the "Modified capacitance analysis" section below). If the source-drain (SD) bias is negligible ($V_{DS} \ll \tilde{\mu}$), we can assume that both source and drain are at zero electro-chemical potential and therefore $e\varphi_{NT} + \tilde{\mu} = 0$. Substituting into Eq. 6, we obtain the relation:

$$V_g = \frac{e(n-p)}{C_g} + \frac{\tilde{\mu}}{e}\frac{1}{\alpha} \qquad \text{(Eq. 8)}$$

We solve Eqs. 8 and 6 self-consistently and substitute into Eqs. 3-5 to calculate I(V$_g$). We find the physical parameters of the device from fitting the theoretical I(V$_g$) to our conductance measurement (Fig. 4b), from which we obtain a band gap of E$_g$=192±5 meV. Then, we substitute the modified band gap $\widetilde{E_g} = E_g \pm \delta E_g$, in which $\delta E_g$ is our upper-bound estimation for the tension induced band gap modulation, into Eq. 6, solve the equations again self-consistently, and obtain how the current should change. Unfortunately, when substituting the above-estimated $\delta E_g$, we anticipate a change of only $|\delta I| = 52\ pA$. A back-of-the-envelope calculation gives $\delta I \approx I_0 \cdot \delta E_g/(2k_BT) \approx 100\ pA$, which agrees well with the more accurate calculation. Meaning, that the change in tension as a result of the snap-through transition can only predict a "jump" in current which is two orders of magnitude smaller than the "jump" measured in the experiment.

**Modulation of the Schottky barrier as a result of strain.** Taking a closer look at Fig. 2, we notice that the negative "jumps" occur only when dI/dVg is small, i.e., when the contact resistance is more dominant than the intrinsic CNT resistance (see Fig. S1 as an example). This indicates that the "jump" might, in fact, be related to the contacts. Since change in strain can enlarge or reduce the band gap, it can also increase or decrease the Schottky barrier (Fig. S2). Assuming thermionic emission, $R_c \propto e^{-\frac{\Phi_B}{k_B T}}$, where $\Phi_B$ is the barrier height, $k_B$ is the Boltzmann constant, and T is the temperature. Since $\delta \Phi_B = \delta E_g \ll E_g$, we can approximate $e^{-\frac{\delta \Phi_B}{k_B T}} \approx 1 + \beta \cdot \delta \Phi_B$ ($\beta = \frac{1}{k_B T}$), and the resulting modulation of the contact resistance should change as $\delta R_c \propto e^{-\beta \Phi_B} \cdot \beta \delta \Phi_B$. Hence, we obtain the relation:

$$\frac{\delta R_c}{R_c} = \beta \cdot \delta E_g \qquad \text{(Eq. 9)}$$

Assuming saturation, $R_{total} \approx R_c$, the expected change in current is thus given by:

$$\delta I = -I \cdot \frac{\delta R_c}{R_c} \qquad \text{(Eq. 10)}$$

Using $|\delta E_g| \approx 1.3242 \cdot 10^{-5} eV$, as estimated above, we receive an anticipated "jump" of only $|\delta I| = I \cdot |\delta E_g| \cdot \beta \approx 30.42 \, pA$. Meaning, that the effect of Schottky barrier modification due to the change in strain can also only predict a "jump" in current which is two orders of magnitude smaller than the "jump" measured in the experiment. Even if we take into account both effects that arise from the band gap modification as a result of strain, they still cannot account for the experimental "jump" in current.

**Modified capacitance analysis.** Since tension cannot account for the "jumps" measured in the experiments, we turn back to the more intuitive explanation, that the "jump" in the current results from the change in capacitance due to the mechanical ST. The above model (Eq. 1) assumes $\frac{\partial I}{\partial (V_g^{DC} C_g)} = \frac{\partial I}{C_g \partial V_g^{DC}}$. However, we know that the capacitance is not constant and changes with the CNT movement. We wish to estimate the change in current due to the partial derivative of the current with respect to the capacitance, i.e., to utilize the following relation:

$$\delta I = \frac{\partial I}{\partial C_g} \cdot \delta C_g \qquad \text{(Eq. 11)}$$

First, since we know the exact CNT shape (z(x)) for every static load, we can calculate the capacitance between the CNT and the local gate for every load according to:

$$C_g(V_g) = \int_0^L \frac{\alpha_g 2\pi\varepsilon_0}{\ln\left(\frac{2(g_0+z(x))}{r}\right)} dx \qquad \text{(Eq. 12)}$$

where $\varepsilon_0$ is the vacuum permittivity. $\alpha_g = \frac{C_g}{C_{total}}$ is the gate efficiency factor[23], and in order to be consistent, we use the same value received from the resonance frequency fit, which is mostly determined by the gate voltage at which the ST transition occurs. The calculated capacitance gate dependence according to Eq. 12 is depicted in Fig. 4a, from which we extract the capacitance difference due to the ST buckling transition: $\delta C_g = 0.64\ aF$.

Next, in order to estimate the partial derivative $\frac{\partial I}{\partial C_g}$, we return to our conductance model (Eqs. 3-8), and substitute the real $C_g(V_g)$ to correct our theoretical fit to the data (Fig. 4b). Then, we add a 3% change in the capacitance, solve the equations once more, and estimate the partial derivative as the difference between the new current and the original current at the same DC voltage at which the "jump" occurs (in our example, $V_{g,ST} = 2.77\ V$): $\frac{\partial I}{\partial C_g} = \frac{I_2(V_{g,ST}) - I_1(V_{g,ST})}{0.03 \cdot C_g(V_{g,ST})}$.

Substituting into Eq. 11 yields an anticipated $\delta I_{V_{g,ST}=2.77V} = 5.8 \pm 0.1\ nA$, which is in good agreement with the measured "jump" in the current, $\delta I_{exp,1} = 4.77\ nA$. We do the same for the "jump" at $V_{g,ST} = -2.42\ V$ and obtain $\delta I_{V_{g,ST}=-2.45V} = 6.2 \pm 0.1\ nA$, also in relatively good agreement with the experimental $\delta I_{exp,2} = 4.35\ nA$ (Fig. 4b).

**Capacitance calculation.** In order to verify our capacitance estimation, we also model our device geometry in COMSOL (Fig. S3). A full electrostatic analysis with the calculated CNT configuration (z(x)) after the transition yields $C_{g,down}^{COMSOL} = 1.083\ aF$, fairly close to our estimated $C_{g,down}^{Eq.12} = 1.2\ aF$. For the upward configuration we obtain: $C_{g,down}^{COMSOL} = 1.765\ aF$, in good agreement to our estimated $C_{g,down}^{Eq.12} = 1.84\ aF$. These results are extremely encouraging. First, they affirm that our estimation of $\alpha_g$ from the resonance fit is reasonable for this device geometry. Second, if we use $\delta C_g^{COMSOL} = 0.682\ aF$, we get a prediction of $\delta I_{V_{g,ST}=2.77V} = 6.26\ nA$, still comparable to the experimental "jump" and consistent with our estimation.

**Unsolved mysteries.** We present the results of a similar analysis from two additional bi-stable devices in the Supplementary Information (Table S1), from which we can also deduce that the modified capacitance model achieves the best prediction. We believe that the change in capacitance is the dominating mechanism affecting the CNT conductance at the ST transition. And indeed, we were able to achieve good predictions for several devices, in which the "jump" occurs at gate voltages for which the CNT intrinsic resistance is most dominant (Table S1). Unfortunately, negative "jumps" remain a puzzle. As stated before, we observe that negative "jumps" occur only when the current is roughly constant with respect to the gate voltage, and the contact resistance is most dominant in setting the total CNT resistance. Therefore, we believe that there is another mechanism, in which the ST transition also affects the contact (either competing or contributing). Further investigation of the data reveals that negative "jumps" were only observed in devices with relatively thick CNT diameters (d~4-6nm, measured in AFM). This implies that these CNTs are likely to have two or more walls. Hence, one possibility for a contact effect could be inner-shell sliding[24,25] relative to the fixed outer shell which adheres to the metallic electrodes, due to stretching. Such sliding is ought to modify the contact resistance, and can either increase or decrease the CNT current. We are currently pursuing this direction and further experiments are needed in order to explore this hypothesis.

**Large hysteresis window.** Most "jumps" are also characterized by noticeable hysteresis, as predicted by ST theory. There is no clear correlation between the "jump" height, $\delta I$, and the hysteresis window size (Fig. S4a), even though both parameters are essentially dependent on the initial CNT configuration. Nevertheless, it is understandable, as the "jump" height also depends on the CNT conductivity, which is independent of the initial configuration. In a few cases, we encounter a device with a very large relative hysteresis window, $\Delta V_{G,hyst}/V_{G,ST} \sim 1$, as in the transfer characteristics in Fig. S4b for example. Latching refers to a nonvolatile trait of a bi-stable buckled beam device, in which the beam remains in the downward configuration after the electrostatic force is removed. Our theory predicts that under certain initial conditions, latching should be realizable in CNT bi-stable devices[27]. The latching criterion is usually formalized by the demand that the force at which a snap-back occurs must be negative ($F_{SB}<0$). However, if we define the relative hysteresis window as $\Delta V_{G,hyst}/V_{G,ST}$, the latching criterion can also be written as $\Delta V_{G,hyst}/V_{G,ST} >1$. In our devices we cannot test this theory, as we only have a single local gate, and therefore cannot apply a repulsive force to observe the snap-back of the device. However, if we consider the hysteresis criterion, we know that it is definitely in reach, as we have observed relative hysteresis windows approaching 1 (orange dots in Fig. S4a).

**Conclusions.** In summary, we present experimental data, signature of bi-stable CNT resonators, in which mechanical ST transition is accompanied by a change in the DC conductance of the device. We attempt to explain the seemingly intuitive phenomenon by several models. We show that the change in strain cannot account for the significant DC "jumps" measured, neither because of tension modulation of the contacts resistance, nor because of modulation of the intrinsic CNT conductance. We conclude that the dominant mechanism causing the "jump" is the change in capacitance, which is not trivial to predict without knowing the exact position and shape of the CNT. Together with a Drude-based model for the CNT conductance, we calculate the contribution of the capacitance change alone to the current, and obtain excellent agreement with the experimental data. However, the capacitive model cannot account for all of the experimental data, and specifically a decrease in the conductance at the ST transition. We attribute this decrease to a change in the contacts, possibly inner shell sliding, but this hypothesis has not been experimentally validated. We believe that the profound understanding of the conductance "jumps" of bi-stable CNT resonators can lead to better and more robust design of such devices, and we present the indication of realistically achievable latching devices as an example.

**Methods**

**Device Fabrication.** All CNT resonators were fabricated according to the local gate self-aligned technique reported in Ref. 26. Briefly, Source and Drain electrodes were patterned using standard e-beam lithography and Cr/Pt 7/32 nm metal layers were evaporated. After the SD patterning, a critical BOE etching step is required[19], after which the local gate was patterned self-aligned, and another layer of Cr/Pt 7/32 nm metals were evaporated. Finally, CNTs were grown in CVD using the fast heating growth technique[28] with H2/CH4 0.5/0.5 SLM flow at 900°C.

**Conductance measurements.** All conductance measurements, from which the data in Fig. 2 was extracted, were conducted in vacuum of P~4e-4 Torr at room temperature. A $V_{DS}$=10 mV bias was applied between the source and drain, and the current was measured using a Stanford SR-570 low noise current preamplifier.

FIGURES.

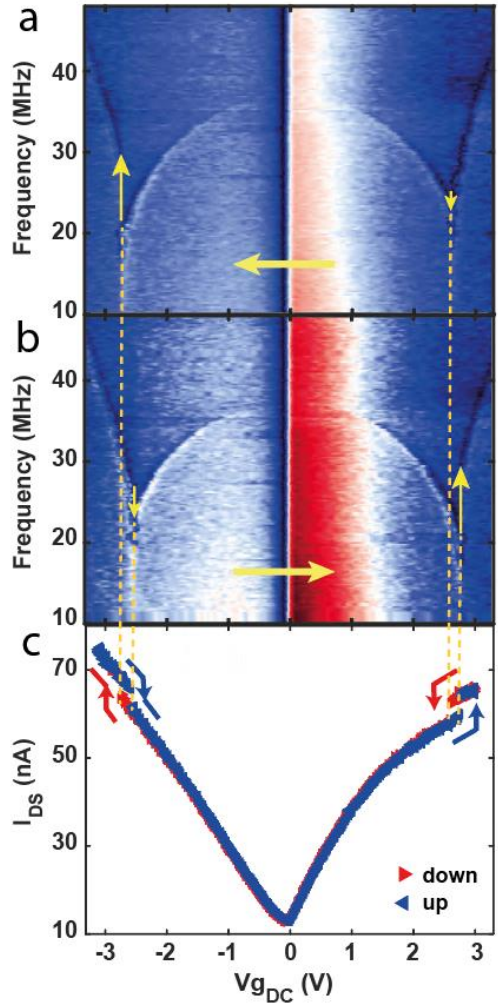

**Figure 1.** Conductance and resonance measurements of a typical bi-stable device. (a-b) Resonance frequency measurement of a typical device exhibiting snap-through bi-stability, consisting of downward (a) and upward (b) gate sweeps, marked by the horizontal yellow arrows. The abrupt transition from upward to downward curvature (and vice versa) is characterized by a "jump" in the resonance frequency, marked by the vertical yellow arrows. (c) Transfer characteristics curve of the same device as in (a-b), exhibiting standard small band gap carbon nanotube characteristics. The ST signature appears as a "jump" in the DC conductance at the same loads as the jumps and hysteresis in (a-b).

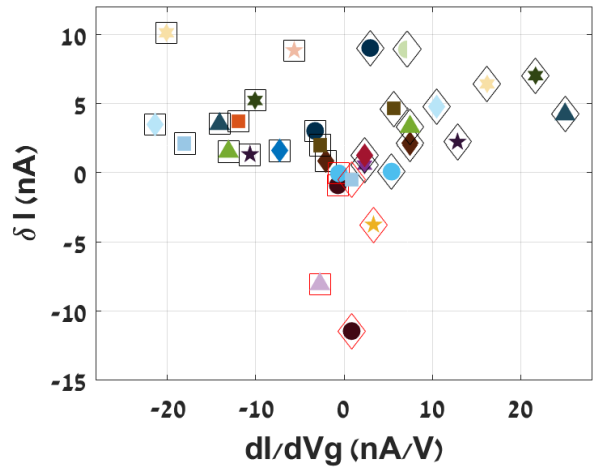

**Figure 2.** δI$_{DS}$ of "jump" vs. $\partial I/\partial V_G$, acquired from 20 different bi-stable CNT resonators. Each device is characterized by marker type and color. The outer frame differentiates between "jumps" for p-type (square) or n-type (diamond) CNTs, and red frames indicate negative "jumps", in which a decrease in the conductance was observed.

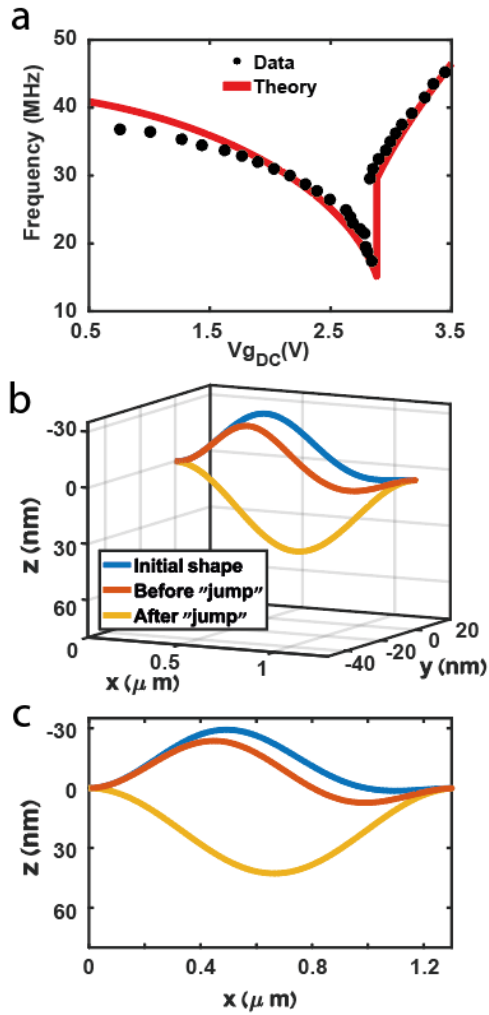

**Figure 3.** (a) Theoretical fit to the resonance data from Fig. 1, according to the theoretical model in Ref. 19. (b) 3D CNT shape at its initial configuration (blue) when no force is applied, and just before (orange) and immediately after (yellow) the ST buckling transition. (c) In-plane component of the CNT configurations presented in (b), the only component affecting its capacitance to the local gate.

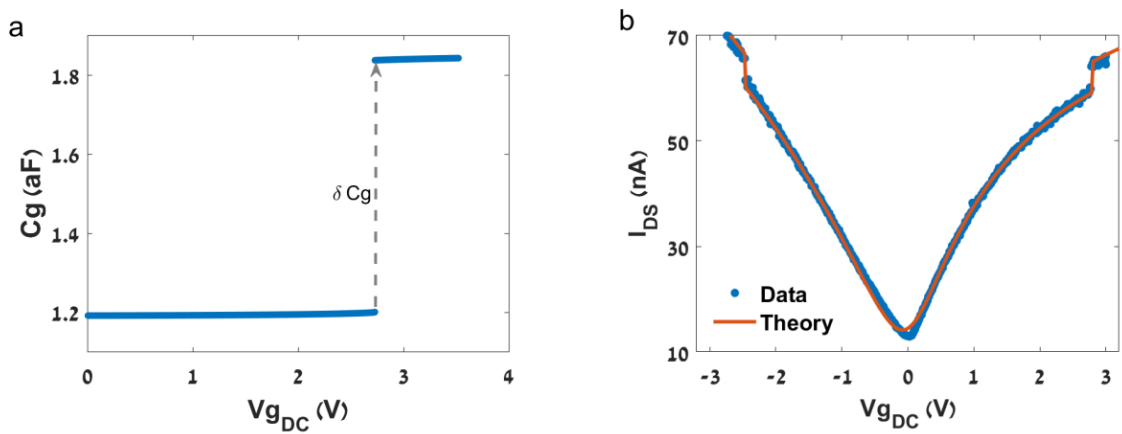

**Figure 4.** Modified capacitance analysis. (a) Gate capacitance as a function of the gate voltage, calculated according to Eq. 12, where z(x) was obtained from the theoretical fit in Fig. 3a for every static load. (b) Theoretical fit according to Eqs. 3-8, where $C_g(V_g)$ was extracted from (a). The theoretical "jump" estimations according to the modified capacitance model were added manually at the ST gate voltages.

TABLES.

| Symbol | Physical parameter | AFM data |
|---|---|---|
| $g_0$ | Height of the source and drain above the local gate | 150 nm |
| r | CNT radius | 1.3 nm |
| L | CNT length | 1.6 μm |

**Table 1.** Physical parameters of the device presented in Fig. 1 (Device I).

ASSOCIATED CONTENT

**Supporting Information**.

The following files are available free of charge.

Supporting Information (PDF)

AUTHOR INFORMATION

**Corresponding Author**

*Yuval E. Yaish, yuvaly@technion.ac.il

**Present Addresses**

†If an author's address is different than the one given in the affiliation line, this information may be included here.

**Author Contributions**

The manuscript was written through contributions of all authors. All authors have given approval to the final version of the manuscript.

**Funding Sources**

This study was supported by the ISF (Grant No. 1854/19), and the Russell Berrie Nanotechnology Institute. S.R. acknowledges support by the Council for Higher Education and the Russel Berrie scholarships.

**Notes**

Any additional relevant notes should be placed here.

ACKNOWLEDGMENT

The work made use of the Micro Nano Fabrication Unit at the Technion.

ABBREVIATIONS

DC, direct current; CNT, carbon nanotube; ST, snap-through

# DC Signature of snap-through bi-stability in carbon nanotube resonators

# Supporting Information

*AUTHORS: Sharon Rechnitz, Tal Tabachnik, Shlomo Shlafman, Michael Shlafman and Yuval E. Yaish[*].*

Andrew and Erna Viterbi Faculty of Electrical and Computer Engineering, Technion, Haifa, Israel.

**Additional Devices**

Following is a table summarizing the analysis in the main text discussion and additional two small band-gap devices exhibiting a positive DC ST "jump" that were analyzed in the same manner as described in the main text:

| Device | $\delta I_{exp}$ Experimental "jump" | Naïve capacitance model | Band gap modulation due to strain | Schottky barrier modulation due to strain | Modified capacitance analysis |
|---|---|---|---|---|---|
| I | 4.77 nA | 1.45 nA | 52 pA | 30.2 pA | 5.8 nA |
| II | 1.3 nA | 0.88 nA | 48 pA | 29.3 pA | 1.6 nA |
| III | 3 nA | 1.38 nA | 80 pA | 62.1 pA | 3.7 nA |

**Table S1.** Results summary of the predicted "jump" from the different theories discussed in the text compared to the experimental "jump" for three different CNT devices.

It can be easily observed that the conclusions presented in the main text for Device I are also valid for devices II and III, and can therefore be generalized.

**Supplementary Figures**

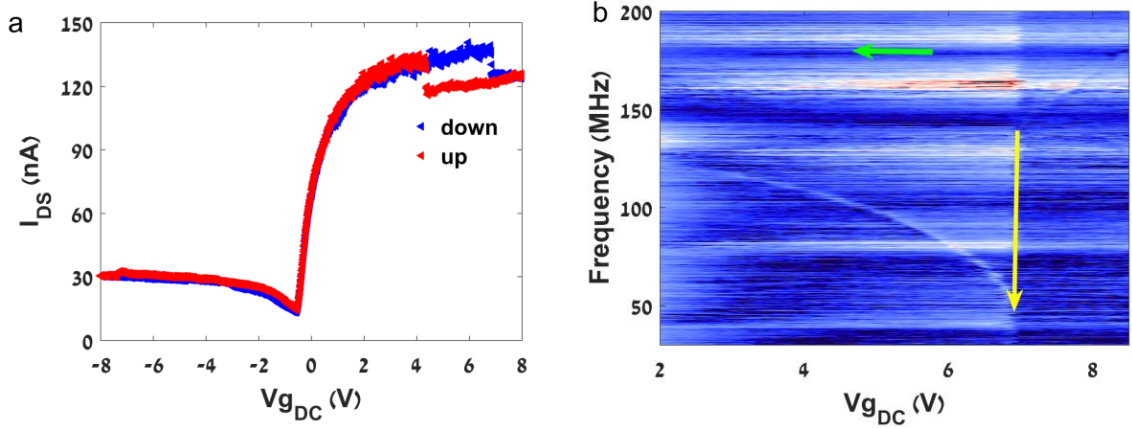

**Figure S1.** (a) Transfer characteristics curve exhibiting a decrease in conductance at the snap-through transition ("negative jump"). (b) Resonance frequency measurement (downward sweep, as indicated by the green arrow) of the same device as in (a) confirming that ST transition (marked by the yellow arrow) occurs at the same gate voltage as the conductance "jump" in (a).

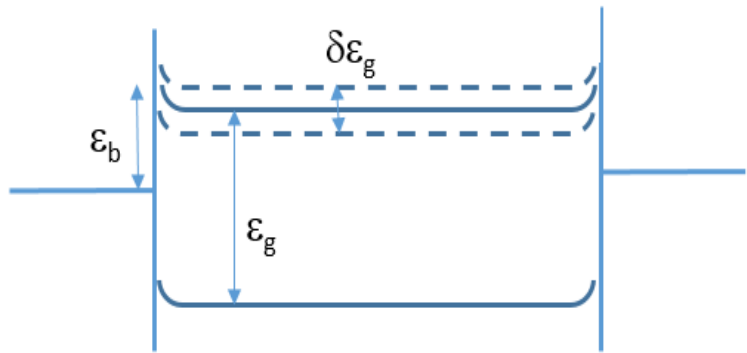

**Figure S2.** Schematic illustration of the Schottky barrier modulation as a result of the band gap modulation due to strain.

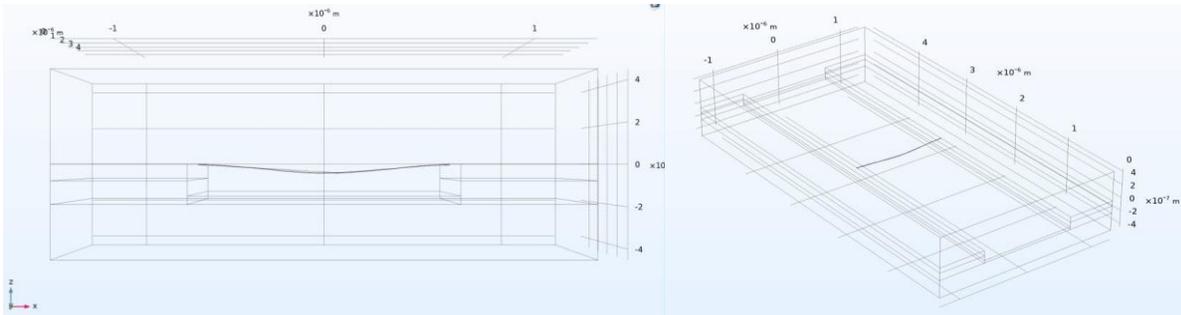

**Figure S3.** Device geometry modelling in COMSOL for capacitance estimation.

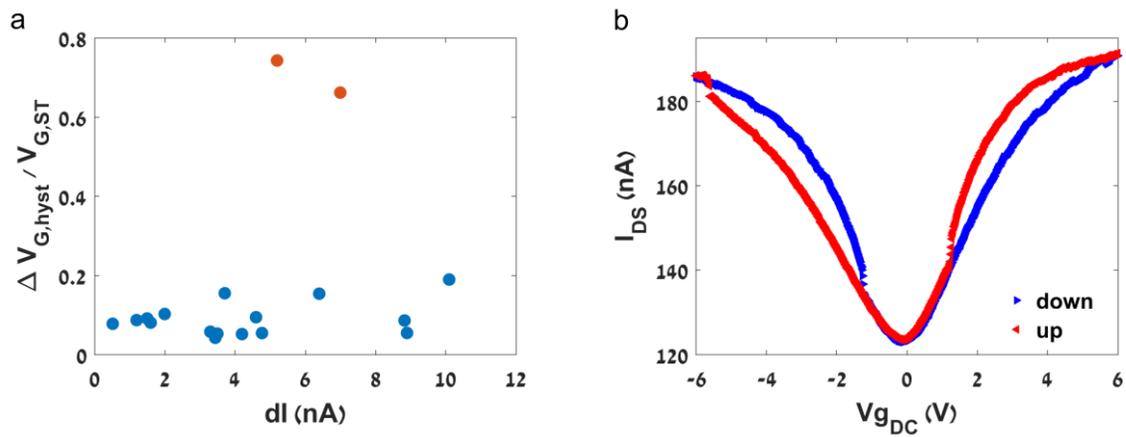

**Figure S4.** (a) Relative hysteresis window as a function of "jump" height. The orange dots represent a relative hysteresis window approaching 1. (b) Conductance measurement of a bi-stable device with a large hysteresis window approaching latching.

4